\documentclass[manuscript]{aastex701}
\let\listofchanges\relax
\usepackage{comment}
\usepackage{xcolor}
\usepackage[normalem]{ulem}
\usepackage[final,commandnameprefix=always]{changes}
\setdeletedmarkup{\textcolor{red}{#1}}

\newcommand{\npfe}{\(\rm {npFe}^{0}\)}
\newcommand{\npFe}{\(\rm {npFe}^{0}\)}
\newcommand{\vphi}{\(\varphi\)}
\newcommand{\um}{\(\mathrm{\mu m}\)}
\newcommand{\sig}{\(\mathrm{\sigma}\)}
\newcommand{\kamo}{Kamo`oalewa}
\newcommand{\Kamo}{Kamo`oalewa}

\begin{document}

\title{Composition and Space Weathering Characteristics of Tianwen-2 Mission's First Target Near-Earth Asteroid (469219) \kamo}

\author[orcid=0009-0001-5533-5671,sname='Liu',gname=Minge]{Minge Liu}
\affiliation{State Key Laboratory of Solar Activity and Space Weather, National Space Science Center, Chinese Academy of Sciences, Beijing, China}
\affiliation{College of Earth and Planetary Sciences, University of Chinese Academy of Science, Beijing, China}
\email{liuminge23@mails.ucas.ac.cn}

\author [orcid=0000-0002-9848-4650,gname=Yazhou, sname='Yang']{Yazhou Yang} 
\affiliation{State Key Laboratory of Solar Activity and Space Weather, National Space Science Center, Chinese Academy of Sciences, Beijing, China}
\email[show]{yangyazhou@nssc.ac.cn}

\author[gname=Yang,sname='Liu']{Yang Liu}
\affiliation{State Key Laboratory of Solar Activity and Space Weather, National Space Science Center, Chinese Academy of Sciences, Beijing, China}
\affiliation{College of Earth and Planetary Sciences, University of Chinese Academy of Science, Beijing, China}
\email[show]{yangliu@nssc.ac.cn}

\author[gname='Jian-Yang',sname=Li]{Jian-Yang Li}
\affiliation{Planetary Environmental and Astrobiological Research Laboratory (PEARL), School of Atmospheric Sciences, Sun Yat-sen University, Zhuhai, China}
\affiliation{Xinjiang Astronomical Observatory, Chinese Academy of Sciences, Urumqi, China}
\email{lijianyang@mail.sysu.edu.cn}

\author[gname=Qing,sname=Zhang]{Qing Zhang}
\affiliation{Planetary Environmental and Astrobiological Research Laboratory (PEARL), School of Atmospheric Sciences, Sun Yat-sen University, Zhuhai, China}
\email{zhangq735@mail.sysu.edu.cn}

\author[gname=Jiang,sname=Zhang]{Jiang Zhang}
\affiliation{Shandong Key Laboratory of Space Environment and Exploration Technology, Institute of Space Sciences, School of Space Science and Technology, Shandong University, Shandong, China.}
\email{zhang_jiang@sdu.edu.cn}

\author[gname=Yongliao,sname=Zou]{Yongliao Zou}
\affiliation{State Key Laboratory of Solar Activity and Space Weather, National Space Science Center, Chinese Academy of Sciences, Beijing, China}
\email{zouyongliao@nssc.ac.cn}

\correspondingauthor{Yazhou Yang, Yang Liu}

\begin{abstract}

The near-Earth asteroid \kamo, a quasi-satellite of the Earth and the target for sample return by China's Tianwen-2 mission, exhibits distinctive spectral characteristics. \chreplaced{This study re-analyzes the visible and near-infrared reflectance spectrum of Kamo`oalewa published by \citet{2021ComEE...2..231S}}{This study analyzes the visible and near-infrared reflectance spectrum of \kamo}, obtained using the Large Binocular Telescope, to infer its mineral composition and space weathering characteristics.
Spectral similarity analysis is performed by comparing the spectrum of Kamo`oalewa to the mean spectra of various types in the Bus-DeMeo taxonomy to make a preliminary constraint on the combined characteristics of surface mineralogy and space weathering effects. To further characterize the mineral composition, a detailed analysis of the 1 \um\ band center is conducted based on spectral data below 1.25 \um\ that have higher signal-to-noise ratios\chdeleted{ and more bands}.
Empirical models for normalized spectra are developed to estimate the Is/FeO content. The results suggest that asteroid \kamo\ has higher olivine abundance than that of typical S-type asteroids and the Moon, exhibiting an immature to submature degree of space weathering. These findings enhance our understanding of the evolution of similar quasi-satellites and provide important implication for the future exploration of Tianwen-2 mission.

\end{abstract}



\section{Introduction}  \label{intro}

The near-Earth asteroid (NEA) (469219) Kamo‘oalewa, provisionally designated as 2016 HO3 and targeted by China's Tianwen-2 asteroid sample-return mission \citep{sktcxb-11-4-lichunlai, 2025SSPMA..55A9501Z}, exhibits properties atypical of most NEAs. Physically, it features a rapid rotation with a period of approximately 28 minutes, a quasi-satellite orbit around Earth, and a small diameter of approximately 50 m (estimated to be less than 100 m) \citep{2021ComEE...2..231S}. Spectroscopically, ground-based observations with the Large Binocular Telescope (LBT) reveal an unusually steep spectral slope—exceeding that of typical S-type asteroids—along with a prominent 1 \um\ absorption band (Band I) \citep{sharkeydata2021,2021ComEE...2..231S}. These distinctive features suggest that \kamo\ may have undergone a unique formation and evolutionary history compared to most S-type NEAs.

Two hypotheses have been proposed for the origin of \kamo: a lunar origin and a main-belt origin. \citet{2021ComEE...2..231S} noted that its reflectance spectrum more closely resembles those of lunar samples than typical S-type asteroids, supporting an origin from lunar ejecta. Orbital dynamics simulations by \citet{2023ComEE...4..372C} demonstrate that lunar ejecta could evolve into \kamo's current quasi-satellite orbit. Subsequent work by \citet{2024NatAs...8..819J} further refined this scenario, suggesting that, if lunar origin, \kamo\ most likely derives from the ejecta of the Giordano Bruno crater on the Moon. \chadded{Likewise, \citet{ZHU2025INN} used orbital dynamics simulations to propose that \kamo\ originates from the ejecta of the Moon's Tycho crater. }In contrast, \citet{2024LPICo3040.1845Z} proposed that \kamo\ is a typical S-type asteroid and has experienced intense space weathering in the main asteroid belt (MB) before migrating to a near-Earth orbit, with its mineral composition being identical to that of ordinary chondrites (OC).

Space weathering is a prevalent exogenic process that modifies the surfaces of airless celestial bodies, primarily resulting from interactions with solar wind ions, micrometeorite impacts, and galactic cosmic rays  \citep{1996M&PS...31..699C,2004AREPS..32..539C,2001JGR...10610039H,2004PhDT........91N,2016JGRE..121.1865P,2019RAA....19...51W}. Space weathering modifies the spectra of airless bodies, adding an extra layer of complexity in the interpretation of the observed spectra of asteroids. 

Assessing the space weathering characteristics of \kamo\ remains challenging due to the highly limited data and its anomalous spectral features. Lunar-style space weathering typically results in simultaneous spectral darkening, reddening, and the weakening of the absorption bands  \citep{2001JGR...10610039H,2016JGRE..121.1865P}. However, only normalized spectra are currently available for \kamo\ \citep{2021ComEE...2..231S}, and its precise size is unknown, precluding evaluations of the overall albedo. Notably, the coexistence of the pronounced spectral reddening with a distinct Band I feature—contrasting with the typical weakening of absorption bands and spectral reddening in lunar-style space weathering—suggests that the space weathering characteristics on this asteroid may be more complex in terms of degree, spectral changing trend and efficiency. 

\chadded{These three aspects of space weathering are described as follows. }The degree refers to the overall maturity level of the asteroid’s surface materials. The trend describes the systematic changes in spectral features induced by space weathering, generally hereafter referred to as the “space weathering spectral trend" for brevity. The efficiency of space weathering refers to the rate at which spectral features, such as the spectral slope, evolve under the influence of space weathering; this efficiency is typically influenced by both the asteroid's composition and its surrounding space environment. The degree and efficiency of space weathering are linked through the asteroid's exposure time to space weathering processes.

The rapid rotation of \kamo\ introduces additional complications. Space weathering typically produces a global fine-grained regolith layer that covers the surface of an airless celestial body, as exemplified by the formation of lunar soil. However, for such a small, fast-rotating asteroid, it is difficult to maintain a stable regolith layer on its global surface. Numerical simulations by \citet{2021Icar..35714249L} and \citet{2024A&A...692A..62R} indicate that cohesion (e.g. van der Waals forces) can retain weathered grains smaller than several centimeters on its partial surface against the centrifugal forces. Yet, its spectrum displays strong reddening, indicative of strong space weathering in the lunar space weathering paradigm as indicated by the high spectral slope.  

In this study, \chreplaced{we provide a new analysis of the visible and near-infrared reflectance spectrum of \kamo\ reported by \citet{2021ComEE...2..231S} in order to}{we analyze the visible and near-infrared reflectance spectrum of \kamo\ to} characterize its surface mineralogy and space weathering characteristics in terms of degree, spectral trend and efficiency. To achieve these objectives, we first employ a spectral similarity analysis by comparing the asteroid's spectrum with the mean spectra of each typical asteroid type in the Bus-DeMeo taxonomy, which is further supported by analyzing the band center of the 1-\um\ absorption feature. Then, we model various spectra (such as of Apollo\chadded{ lunar} samples, S-type asteroids and laboratory laser-irradiated olivine) based on empirical models to determine the space weathering characteristics of \kamo. Finally, its space weathering degree coupled with mineralogical composition are used to explain its rather atypical spectral appearance. 

\section{Data and Methods} \label{sec:DM}
\subsection{Spectral Similarity and Band Center Analysis}

The spectrum of \kamo\ exhibits a steep spectral slope and a prominent absorption feature at 1 \um, with limited data points available near the 2 \um\ absorption band \citep{sharkeydata2021,2021ComEE...2..231S}. \chreplaced{The available spectral curve}{The spectrum} of Kamo`oalewa consists of two components: a visible-near-infrared (VNIR) spectrum with a reasonable signal-to-noise ratio up to 1.25 \um\ and a spectrophotometric measurement using the z, J, H, and K broadband filters \citep{sharkeydata2021,2021ComEE...2..231S}.

We performed spectral similarity analysis between the spectrum of \kamo\ and the typical spectra of various types of asteroids. Totally 12 types of spectrally-classified asteroids (A, L, K, O, Q, R, S, Sa, Sq, Sr, Sv, and V) exhibiting a prominent Band I absorption feature were selected from a total of 25 types in the Bus-DeMeo asteroid spectral taxonomy \chreplaced{\citep{2009Icar..202..160D}}{ \citep{1999PhDT........50B,2009Icar..202..160D,2019Icar..324...41B,BDmeanspecdb}}. We excluded asteroid types lacking a prominent Band I feature to avoid artificially high similarity scores driven primarily by coincidentally slope matching (e.g., D-type asteroids) \citep{2025ApJ...979L...8K}.

The Bus-DeMeo classification system is the prevailing framework for asteroid spectral taxonomy. These types provide insights into the combined characteristics of mineral composition and space weathering effects, with established correlations to specific meteorite classes. For instance, the sequence of spectral types Q → Sq → S → Sr → Sv aligns closely with the OC series LL → L → H. Although the Q → Sq → S transition might reflect a spectral trend induced by space weathering  \citep{2010Natur.463..331B,2022Icar..38014971D}, alternative studies (e.g., \citealp{2019PASJ...71..103H}) argue that Q-type asteroids may have non-fresh, weathered surfaces lacking fine particles, as coarse weathered ordinary chondritic particles can reproduce Q-type spectra.

Considering the relatively large uncertainties within the z, J, H, and K bands, we also conducted similarity analysis only using the data below 1.25 \um. To reduce the weight of spectral slope caused by space weathering or data uncertainty, we also performed comparative analysis on the continuum-removed spectra. 
\chadded{The continuum was determined following the method of \citet{2020Icar..35013901M} by fitting an upper envelope of straight-line segments that lies above or tangent to the reflectance spectrum at selected points, ensuring that the continuum value is greater than or equal to the reflectance at every wavelength.}
By identifying the spectral type that is most similar to Kamo`oalewa’s spectrum, we can make a preliminary constraint on its surface mineralogy and space weathering characteristics. To further characterize the mineral composition, we conducted a detailed analysis on the 1 \um\ band center (BC1) based on spectral data below 1.25 \um\ that have higher signal-to-noise ratios and more bands. For details on the similarity analysis and band center extraction method, please refer to Appendix \ref{ap:CPRMSmethod}.


\subsection{Empirical Space Weathering Model} \label{subsec:SWmethod}

Given that only a normalized spectrum is available for \kamo, most established space weathering models—which rely on absolute reflectance—cannot be applied. To address this limitation, we adapted a model based on the conventional R950/R750 versus R750 space weathering maturity framework \citep{1995Sci...268.1150L,2000JGR...10520377L}, where R750 denotes the reflectance at 750 nm, and R950/R750 is the ratio of the reflectances at 950 nm and 750 nm. 

The works described below were implemented to adapt the conventional model. We compiled spectra from lunar samples in the Lunar Soil Characterization Consortium (LSCC; \citealp{2006Icar..184...83P,2001JGR...10627985T}) and RELAB \citep{2012Icar..220...51B,RELABdb} databases, along with their corresponding FeO contents (wt.\%) and Is/FeO values, to generate a scatter plot. The x-axis in the plot represents the R750/R550 ratio of each spectrum, which correlates positively with the continuum slope; the y-axis represents the R950/R750 ratio, which correlates negatively with the depth of the Band I absorption feature. This approach ensures that all spectra in the scatter plot are effectively normalized to unity at 550 nm, regardless of whether the original data are absolute or normalized reflectance. As lunar-style space weathering progresses, the spectrum reddens and the absorption features weaken, resulting in a systematic pattern that shifts data points from the lower left to the upper right of the scatter plot. According to \citet{2010Icar..209..564G}, the S-type asteroids Ida and Eros exhibit distinct space weathering spectral trends. Consequently, to validate the capability of the scatter plot to reflect space weathering spectral trends, we incorporated reflectance data from  two asteroids: Ida, acquired by the Solid-State Imaging (SSI) multispectral camera during the Galileo mission \citep{1992SSRv...60..413B,GOIDADB}; and Eros, obtained by the Multi-Spectral Imager (MSI) on the Near-Earth Asteroid Rendezvous (NEAR) mission \citep{1997SSRv...82...31H,NEAREROSDB}. Additionally, we incorporated the results of laser irradiation experiments on olivine from \citet{2017A&A...597A..50Y}, which simulate space weathering effects, to analyze the possible manner in which the scatter plot reflects space weathering efficiency. Detailed descriptions on the traditional R950/R750 versus R750 model and our modified R950/R750 versus R750/R550 model are provided in Appendix \ref{ap:B}.

The modified model revealed a strong correlation between the distance of each point from the origin (0, 0)—defined here as NOMAT (normalized optical maturity, Equation \ref{eq:A3})—and its corresponding Is/FeO value. Is/FeO is a commonly used parameter to quantify space weathering, representing the ratio of ferromagnetic resonance intensity (Is) to the total iron content (FeO) of the sample \citep{1978LPSC....9.2287M}. Between \(\ln{(Is/FeO)}\) and NOMAT, the Pearson correlation coefficient is 0.85, indicating a linear relationship, while the Spearman correlation coefficient is 0.83, suggesting a monotonic association. Both coefficients exceed 0.8, confirming a robust correlation. 
Accordingly, we fitted an exponential model relating NOMAT to Is/FeO using lunar sample data to estimate the Is/FeO value for \kamo\ and thereby infer its degree of space weathering.

To validate our Is/FeO versus NOMAT model, we employed a two-step approach. First, we utilized the measured spectra of Chang'e-5 (CE-5) samples \citep{2022NatCo..13.3119L} to compute the range of Is/FeO values predicted by our model and compared this range with the corresponding measured Is/FeO values \citep{2024Icar..41015892Q}. Second, we independently estimated the Is/FeO values for \kamo\ using an alternative method and compared these estimates with the model-predicted range. Specifically, this method involved calculating the nano-phase iron (\npfe) volume fraction for \kamo\ based on the model proposed by \citet{2006Natur.443...56H}, followed by applying the empirical relation established by \citet{1980LPSC...11.1697M} to convert the \npfe\ fraction to Is/FeO value. Details of the validation methods, including the CE-5 sample data, Hiroi model, and Morris model, are presented in Appendix \ref{ap:C}.

\section{Results}
\subsection{Spectral Similarity and Band Center Analysis}

\begin{figure*}[ht!]
\plotone{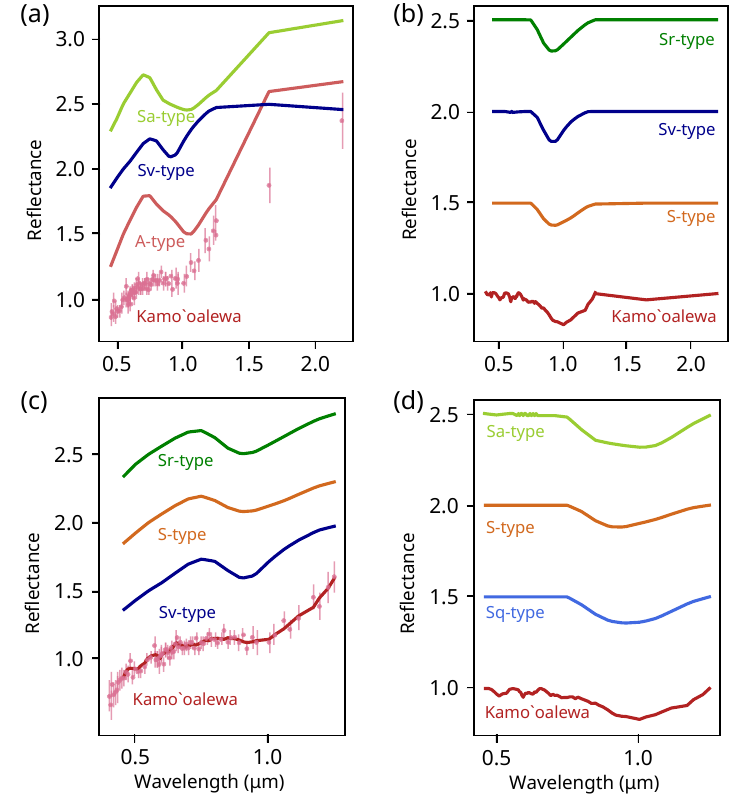}
\caption{Spectral comparison of asteroid \kamo\ with the mean spectra of spectral types in Bus-DeMeo taxonomy \citep{2009Icar..202..160D,BDmeanspecdb} exhibiting prominent Band I absorption feature. (a) Full-band comparison with continuum retained, all spectra normalized at 0.55 \um. (b) Full-band comparison after continuum removal based on normalized spectra. (c) VNIR comparison with continuum retained, all spectra normalized at 0.55 \um. (d) VNIR comparison after continuum removal based on normalized spectra. In each panel, \kamo's spectrum is shown at the bottom, with other spectra ranked by similarity from highest to lowest (bottom to top) and offset vertically for clarity. Panels (a) and (b) employ similarity analysis weighted by wavelength intervals and reflectance errors, whereas panels (c) and (d) use unweighted analysis due to more uniform data point distribution. Data points with error bars in (a) and (c) represent \kamo's original measurements, while its spectral curves in (c) and (d) are smoothed using a Savitzky-Golay filter (window length 11, polynomial order 3; \citealp{SGFM}). \chadded{The smoothed full-band spectrum of \kamo\ in panel (b) consists of the smoothed VNIR spectrum and the original H and K data points.}}
\label{fig:1}
\end{figure*}

We compared the reflectance spectrum of \kamo\ with the mean spectra of the twelve Bus-DeMeo taxonomic types that exhibit a prominent Band I feature. The continuum-removed spectrum of \kamo, which focuses on its mineral composition, is most similar to the mean spectrum of S-type. On the other hand, the spectrum without continuum removal, which can reflect space weathering spectral trend to some extent, aligns most closely with the mean spectrum of A-type. These spectral comparison results, including both full-band and VNIR spectra for the continuum-removed and non-removed cases, are illustrated in Figure \ref{fig:1}. Spectral similarity analyses using full-band spectra (Figure \ref{fig:1}a and \ref{fig:1}b) and VNIR spectra (Figure \ref{fig:1}c and \ref{fig:1}d) yielded similar results: among all the spectral types shown in Figure \ref{fig:1} as being similar to Kamo`oalewa, apart from the A-type (which exhibits the highest spectral slope), the others all belong to the S-complex.

\begin{figure*}[ht!]
\plotone{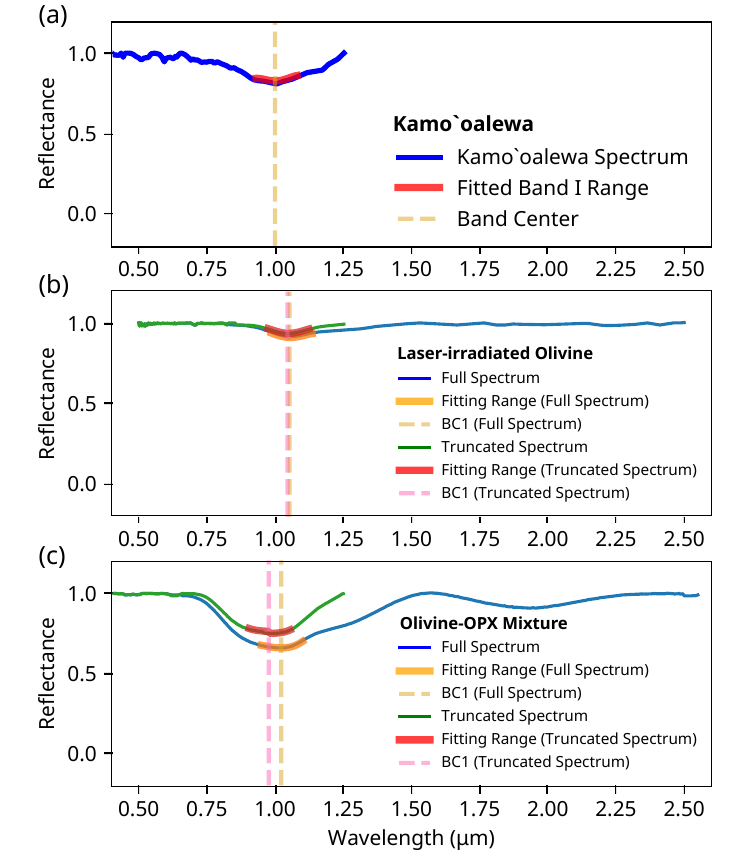}
\caption{BC1 analysis of spectra from different samples. (a) Spectrum of \kamo; (b) \chreplaced{spectrum}{spectra} of laser-irradiated olivine (25 mJ, 4 pulses\chadded{; \citealp{2017A&A...597A..50Y}}); (c) \chreplaced{spectrum}{spectra} of olivine and OPX mixtures (90 wt.\% olivine, 10 wt.\% OPX\chadded{; \citealp{RELABdb}}). The spectrum in (a) extends only to 1.25 \um, with the BC1 fitted using the method of \citet{2014Icar..234..132H} being 1.00 \um. In (b) and (c), both truncated spectra up to 1.25 \um\ and full-wavelength spectra were used. Truncating the spectrum at 1.25 \um\ causes the calculated BC1 to shift toward shorter wavelengths; thus, \kamo's actual BC1 is at least 1.00 \um\ or (more likely) longer, suggesting a high olivine content. The fitting range refers to the region around the absorption band center fitted with a fourth-order polynomial, following the method of \citet{2014Icar..234..132H}.}
\label{fig:15}
\end{figure*}

We performed continuum removal \chadded{\citep{2020Icar..35013901M} }and 1 \um\ band center fittings \chadded{\citep{2014Icar..234..132H} }using only the portion of \Kamo's VNIR spectrum at wavelengths shorter than 1.25 \um. The BC1 value obtained was 1.00 \um\ (Figure \ref{fig:15}a). Furthermore, after performing BC1 calculations on laser-irradiated olivine spectra \citep{2017A&A...597A..50Y}, as well as the spectra of McCoy's olivine and orthopyrexene (OPX) mixtures (Figure \ref{fig:15}b and \ref{fig:15}c; \citealp{RELABdb}), we found that, for laser-irradiated olivine and mixtures with high olivine content, the BC1 value calculated using only wavelengths shorter than 1.25 \um\ of the spectrum was biased toward shorter wavelengths relative to that calculated using the full-wavelength spectrum (Figure \ref{fig:15}b and \ref{fig:15}c).

\subsection{Estimation of the Degree of Space Weathering of \kamo}
\subsubsection{Estimation Using the Normalized OMAT Model}

\begin{figure*}[ht!]
\plotone{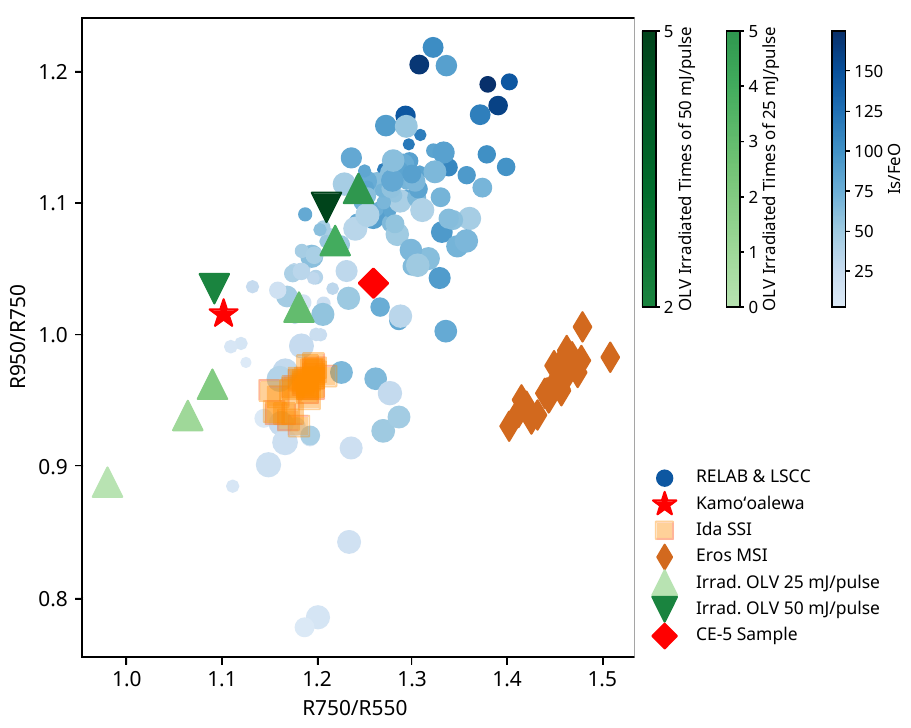}
\caption{Scatter plot of R950/R750 versus R750/R550 for various sample spectra. The blue data points represent LSCC and RELAB lunar samples \citep{2006Icar..184...83P,2012Icar..220...51B}, exhibiting a progression from the lower left to the upper right, which corresponds to increasing Is/FeO values; point size indicates the corresponding FeO content. No variation associated with FeO content is observed. These blue points define the range of spectral changing trend induced by lunar-style space weathering, with the \kamo\ data point positioned at the upper left edge of this range. Refer to the legend for the interpretation of the remaining data points.}
\label{fig:2}
\end{figure*}

Figure \ref{fig:2} presents the scatter plot of R950/R750 versus R750/R550 for various samples and objects. The data points for the LSCC and RELAB lunar samples \citep{2006Icar..184...83P,2012Icar..220...51B} define the range of spectral  changing trend induced by lunar-style space weathering, exhibiting a clear increase in Is/FeO from the lower left to the upper right. The data points for Eros deviate completely from the lunar trend, whereas Ida aligns with this trend, consistent with the summary in \citet{2016JGRE..121.1865P}. The points for \kamo\ and laser-irradiated olivine \citep{2017A&A...597A..50Y} lie on the edge of the range for lunar samples.

\begin{figure*}[ht!]
\plotone{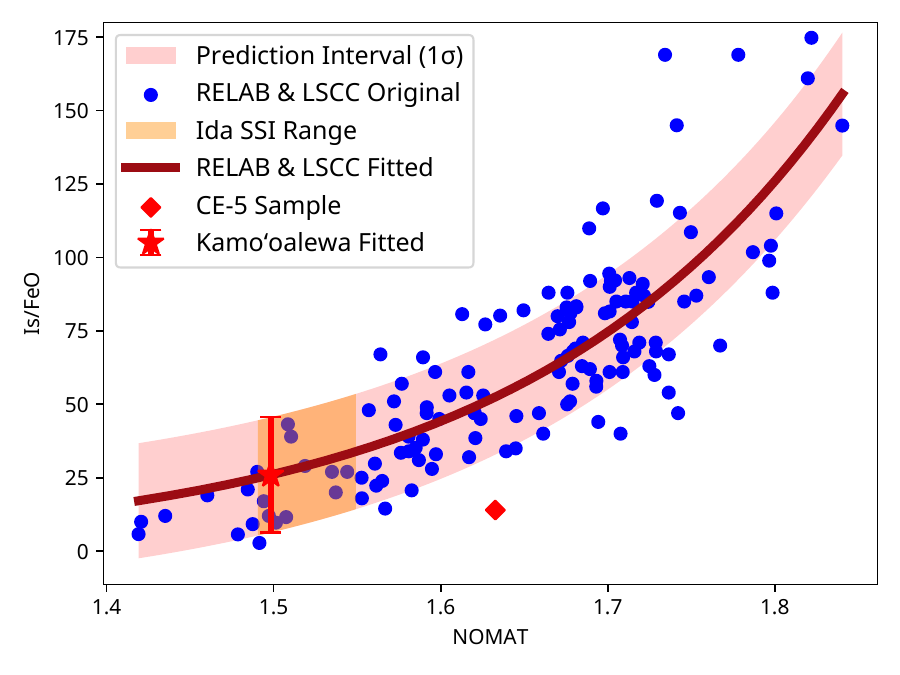}
\caption{Empirical model of Is/FeO versus NOMAT, fitted using RELAB and LSCC lunar sample data. Blue dots represent the original data points used for the fit. The red shaded region indicates the 1\sig\ prediction interval of the model. Data points from the Chang’e-5 sample lie outside the 1\sig\ prediction interval but within the 2\sig\ prediction interval. The orange shaded region denotes the Is/FeO value range for S-type asteroid Ida. The fitted Is/FeO value for \kamo\ is 25.97 ± 19.62.}
\label{fig:3}
\end{figure*}

The NOMAT was calculated for each point based on its coordinates in Figure \ref{fig:2} (see Section \ref{subsec:SWmethod} and Appendix \ref{ap:B} for detail). The exponential fitting relationship between NOMAT and Is/FeO is given by Equation (\ref{eq:1}) (Figure \ref{fig:3}).

\begin{equation}
Is/FeO=0.01022\times e^{5.23280\times NOMAT} \label{eq:1}
\end{equation}

The calculated NOMAT value for \kamo\ is 1.50, corresponding to an Is/FeO value of 25.97±19.62 (1\sig\ prediction interval) according to Equation (\ref{eq:1}). When compared to the lunar space weathering maturity system \citep{1978LPSC....9.2287M}, the value falls within the immature to submature range.

\subsubsection{Validation of Estimated Results}

The position of the CE-5 sample in the R750/R550 versus R950/R750 scatter plot falls within the range of lunar-style space weathering spectral trend (Figure \ref{fig:2}). Additionally, the measured Is/FeO values (Appendix \ref{ap:c1}) reported by \citet{2024Icar..41015892Q} lie within the 2\sig\ prediction interval (but outside the 1\sig\ prediction interval) of our fitted model (Figure \ref{fig:3}). 

The \npfe\ contents in \kamo, calculated using the Hiroi model, are 0.095 vol.\% and 0.104 vol.\% for the two cases that \kamo\ evolves from ordinary chondrites (OC) and lunar mare ejecta, respectively. The corresponding Is/FeO values, estimated using the Morris model \citep{1980LPSC...11.1697M}, are 43.92, and 45.53, both within the 1\sig\ prediction interval of our model (See Appendix \ref{ap:C} for detail). These results demonstrate the credibility of our model fitting outcomes and provide insights for the subsequent discussions on the mineral composition of \kamo.

\section{Discussions}

\subsection{Mineral Composition of \kamo}

\citet{2021ComEE...2..231S} and \citet{2021Icar..35714249L} suggested that \kamo\ may be classified as an S-type asteroid. Due to the lack of coverage for the 2 \um\ band in the available data, it remains challenging at the present to decisively classify its spectral type. Based on our quantitative similarity analysis using the continuum-removed spectrum, \kamo\ is most similar to S-type asteroids. Considering that the Bus-DeMeo taxonomy does not consider spectral slope as a primary classification criterion, particularly for the S-complex asteroids \citep{2009Icar..202..160D}, we thus classify \kamo\ as an S-type asteroid, which aligns with the analysis by \citet{2024LPICo3040.1845Z} and consistent with the interpretation of \citet{2021ComEE...2..231S}. Once higher-quality spectral data are acquired, the method proposed by \citet{2022A&A...665A..26M} can be used to obtain more accurate spectral classification results.

The surfaces of S-type asteroids are primarily composed of silicate minerals, such as olivine and pyroxene, along with a small amount of iron-nickel metals. Therefore, the dominant minerals on the surface of \kamo\ should be olivine and pyroxene. Our spectral similarity analysis using the normalized spectrum of \kamo\ without continuum removal indicates that its spectrum is most similar to the average spectrum of A-type asteroids, which are typically rich in olivine \citep{2019Icar..322...13D}. Combining the above evidence, we conclude that \kamo\ likely contains a higher olivine content than typical S-type asteroids.

This conclusion is further supported by the BC1 analysis. Truncation of the spectra at 1.25 \um\ may lead to an apparent shift of the band center toward shorter wavelengths compared to the full spectra (Figure \ref{fig:15}b and \ref{fig:15}c), particularly for samples with high olivine proportions (exceeding 50 wt.\%). This artifact arises because the truncation renders the right half of the Band I feature incomplete, biasing the continuum low overall, with greater bias toward longer wavelengths. Additionally, since 1.25 \um\ is near the M1-2 feature of olivine \citep{burns1970am,1987JGR....9211457K,2015Icar..252...39C,2022Icar..37314765P}, this truncation makes the M1-2 feature nearly indiscernible. Our calculations indicate that the center of Kamo`oalewa's Band I, as calculated from the truncated spectrum, is 1.00 \um, implying that its actual Band I center must be at least 1 \um\ or longer. According to \citet{1993Icar..106..573G}, this suggests a high olivine content relative to typical S-type asteroids.

Additionally, the fitted results from our model show a low Is/FeO value for \kamo, while those from Hiroi model (Appendix \ref{ap:c2}) indicate a high \npfe\ content (in both hypothetical cases of \kamo\ origin from main belt or typical lunar mare ejecta). According to the Morris model (Appendix \ref{ap:c3} and Equation \ref{eq:a6}), this suggests a high FeO content for the surface materials of \kamo. 

\subsection{Space Weathering Characteristics of \kamo}

We examine the space weathering characteristics on \kamo\ from three perspectives: the degree, spectral trend, and efficiency of space weathering.

Regarding the degree of space weathering on \kamo, the Is/FeO value of \kamo\ derived from our Is/FeO versus NOMAT model (Figure \ref{fig:3}) indicates an immature to submature state when compared with lunar samples \citep{1978LPSC....9.2287M}. Although \kamo\ exhibits notably reddened spectral slope, it is inadvisable to infer a high degree of space weathering solely from this feature, as the 1 \um\ absorption band in its spectrum remains relatively prominent. Our findings align with the asteroid's extremely rapid rotation rate and small diameter, which collectively inhibit the development of global and stable regolith layer on its surface \citep{2024A&A...692A..62R}. 

Regarding the spectral changing trend induced by space weathering of \kamo, the continuum-preserved spectrum exhibits the highest similarity to the mean spectrum of A-type asteroids. This similarity might reveal that the space weathering spectral trend of \kamo\ is expected to align most closely with that of pure olivine minerals or A-type asteroids. For a more comprehensive understanding of this trend, the scatter plot of R950/R750 versus R750/R550 (Figure \ref{fig:2}) provides valuable insights. As shown in Figure \ref{fig:2}, the data point for \kamo\ lies at the upper-left margin of the envelope defined by the lunar-style space weathering spectral trend, indicating consistency with the lunar-style trend. However, its near coincidence with the point for laser-irradiated olivine suggests that the space weathering spectral trend of \kamo\ might more closely matches that of olivine or A-type asteroids. Furthermore, in this plot, the distributions of the S-type asteroids Ida and Eros indicate that Ida follows the lunar-style space weathering spectral trend, whereas Eros deviates significantly from it. This is consistent with the report by \citet{2004AREPS..32..539C} and \citet{2016JGRE..121.1865P}, thereby lending credence to our application of this methodology in assessing the space weathering spectral trend for \kamo.

In the scatter plot of R950/R750 versus R750/R550, olivine samples irradiated with a pulse energy of 50 mJ/pulse plot systematically above and to the left of those irradiated with 25 mJ/pulse (green triangles in Figure \ref{fig:2}; \citealp{2017A&A...597A..50Y}). This indicates that, in the scatter plot, in addition to the primary progression of increasing space weathering maturity from the lower left to the upper right, a secondary pattern of escalating space weathering efficiency may exist from the lower right to the upper left. Consequently, with its data point located at the upper-left edge of the range of lunar data points, the position of \kamo\ in the plot may indicate a relatively high space weathering efficiency.

\subsection{Connecting Mineral Composition to Space Weathering Characteristics}

Investigations of the space weathering characteristics and mineralogical composition of \kamo\ are mutually informative, because the spectral responses to weathering are intrinsically mineral-dependent. For instance, olivine exhibits approximately five times greater spectral slope enhancement than orthopyroxene under the same space weathering conditions \citep{2002AdSpR..29..783S,2009Natur.458..993V}. The spectrum of \kamo\ exhibits both a very high slope and a distinct 1 \um\ absorption feature, which appears contradictory upon initial inspection. However, this apparent contradiction underscores the self-consistency of our interpretation regarding the asteroid's high olivine content and low degree of space weathering. Specifically, the low degree of space weathering preserves the distinct 1 \um\ absorption feature, while the high olivine content allows the spectrum to maintain a steep slope even under minimal space weathering.

More specifically, the observations by \citet{2009Natur.458..993V} reveal that certain young silicate asteroid families exhibit spectral slopes comparable to or even exceeding those of older asteroid families. This phenomenon is attributed to the higher olivine contents in these young families, which enhances the efficiency of spectral reddening induced by space weathering \citep{2009Natur.458..993V}. After corrections for mineral composition, \citet{2009Natur.458..993V} inferred that the degrees of space weathering for those young asteroids are lower than those of older families. Based on the discussions in this paper, we find that the properties of \kamo\ are in good agreement with those of these young asteroids.

\section{Conclusion}

Our analysis indicates that the mineral composition of NEA \kamo\ is consistent with those of S-type asteroids, featuring both a higher olivine content and probably elevated FeO abundance relative to typical S-type asteroids. Space weathering on \kamo\ appears to be at an immature to submature stage, exhibiting a spectral trend consistent with lunar-style space weathering but more closely resembling that of pure olivine minerals or A-type asteroids, meanwhile suggesting a potent space weathering efficiency.

Due to the limited observational data currently available for \kamo, this study leverages the limited available information to provide preliminary references that may inform the implementation of the Tianwen-2 mission, while anticipating comprehensive data from this endeavor.

\begin{acknowledgments}
This work was supported by the National Natural Science Foundation of China (Grant No. 42374218 and 42441804), Beijing Nova Program, and the Climbing Program of National Space Science Center, Chinese Academy of Sciences (E1PD3001). J.-Y. Li's work was partially supported by the 2024 Xinjiang Uygur Autonomous Region Tianchi Talent Program. We thank B. Sharkey for publishing the LBT spectrum of \kamo\ (\citealp{sharkeydata2021}, \href{https://doi.org/10.5281/zenodo.5542337}{https://doi.org/10.5281/zenodo.5542337}). We thank RELAB (\citealp{RELABdb}, \href{https://sites.brown.edu/relab/}{https://sites.brown.edu/relab/}) and LSCC (\href{https://sites.brown.edu/relab/lscc/}{https://sites.brown.edu/relab/lscc/}) databases for providing spectra and compositional data of mixtures and lunar samples. We thank NASA's Planetary Data System, Small Bodies Node (\href{https://pds-smallbodies.astro.umd.edu/}{https://pds-smallbodies.astro.umd.edu/}) for providing hyperspectral image of Ida (Data Product GO-A-SSI-3-IDA-CALIMAGES-V1.0, \citealp{GOIDADB}) and Eros (Data Product NEAR-A-MSI-5-DIM-EROS/ORBIT-V1.0, \citealp{NEAREROSDB}), as well as the mean spectra of Bus-DeMeo Taxonomy (Data Product EAR-A-VARGBDET-5-BUSDEMEOTAX-V1.0, \citealp{BDmeanspecdb}). The data used in this study are deposited in the Science Data Bank (ScienceDB), and can be accessed by doi link: \href{https://doi.org/10.57760/sciencedb.space.03208}{https://doi.org/10.57760/sciencedb.space.03208}.
\end{acknowledgments}

\begin{contribution}
\noindent Conceptualization: Minge Liu, Yazhou Yang, Yang Liu, Jian-Yang Li, Jiang Zhang.

\noindent Data curation: Minge Liu, Yazhou Yang, Yang Liu.

\noindent Investigation: Minge Liu, Yazhou Yang.

\noindent Methodology: Minge Liu, Yazhou Yang, Yang Liu, Jian-Yang Li, Qing Zhang.

\noindent Funding acquisition: Yazhou Yang, Yang Liu, Jian-Yang Li.

\noindent Supervision: Yazhou Yang, Yang Liu, Jian-Yang Li, Yongliao Zou.

\noindent Writing – original draft: Minge Liu, Yazhou Yang, Yang Liu, Jian-Yang Li.

\noindent Writing – review and editing: Minge Liu, Yazhou Yang, Yang Liu, Jian-Yang Li, Qing Zhang, Jiang Zhang, Yongliao Zou.


\end{contribution}

%



\appendix

\section{Similarity Quantification and Band Center Extraction Method} \label{ap:CPRMSmethod}

\subsection{CPRMS Index} \label{ap:AA1}

To quantify spectral similarity, we employed the weighted Centered Pattern Root Mean Square (CPRMS) index \citep{2001JGR...106.7183T,2015GeoRL..42.6945Z}. This index simultaneously integrates differences in spectral shape \citep{2001JGR...106.7183T}. Weights were assigned to account for the non-uniform wavelength distribution of \kamo\ spectral data points and reflectance measurement uncertainties. Analyses were conducted on the spectra both with and without continuum removal \citep{2020Icar..35013901M}, which was applied consistently to the comparison mean spectra of the 12 types. Continuum removal emphasizes the Band I absorption shape, thereby reducing the influence of space weathering on spectral matching and highlighting intrinsic compositional features.

\chadded{The continuum removal followed the method of \citet{2020Icar..35013901M}. The continuum is formed as a piecewise linear envelope that remains at or above the reflectance spectrum at key points, guaranteeing that continuum values are no less than reflectance values across all wavelengths. This approach identifies local maxima or inflection points that define the overall spectral shape while avoiding absorption features.}

\chadded{The procedure for constructing the continuum begins with the first data point as the initial anchor. A trial line segment is formed by connecting this anchor to the next candidate point and extending it to the last data point. If the extended line lies above or on the spectrum across the entire range, the candidate is accepted as the next continuum point, and it becomes the new anchor. Otherwise, the anchor remains fixed, and the process advances to the subsequent candidate point, repeating until a valid segment is found. This iterative selection continues until the end of the spectrum is reached.}

The CPRMS spectral similarity index \citep{2001JGR...106.7183T,2015GeoRL..42.6945Z} is a comprehensive and widely used index that simultaneously accounts for similarities in spectral shape and Euclidean distance. The standard CPRMS formula is given by Equation (\ref{eq:A1}):

\begin{equation}
    CPRMS=\sqrt{\frac{1}{N}\sum_{i=1}^N[(f_i-\bar{f})-(r_i-\bar{r})]^2} \label{eq:A1}
\end{equation}

In this equation, \(f_i\) and \(r_i\) denote the reflectance values at the i-th data point for the \kamo\ spectrum (or the reference spectrum) and the comparison spectrum, respectively; \(\bar{f}\) and \(\bar{r}\) represent the mean reflectance values of the respective spectra. A smaller CPRMS value indicates greater similarity between the two spectra. Notably, CPRMS calculations require that the wavelengths of the reference and comparison spectra align; thus, we resample the comparison spectrum to match the wavelengths of the \kamo\ spectrum. Additionally, since the \kamo\ spectrum is normalized at 0.55 \um\ to maintain consistency with conventional practice, the comparison spectrum is similarly normalized at this wavelength.

Given the uneven wavelength distribution in the \kamo\ spectrum (full-bands) and the substantial variability in reflectance measurement errors across data points, direct application of Equation (\ref{eq:A1}) may inadequately constrain similarities in regions with sparse data points or higher measurement errors. To address this, we incorporate weights for each data point based on wavelength density and reflectance measurement error, assigning higher weights to points in sparse wavelength regions and lower weights to those with larger errors. This approach ensures that the similarity calculation comprehensively captures the features of all data points across the spectrum. The weighted CPRMS formula is expressed as Equation (\ref{eq:A2}):

\begin{equation}
    CPRMS=\sqrt{\frac{1}{N}\sum_{i=1}^{N}{w_i\left[\left(f_i-\bar{f}\right)-\left(r_i-\bar{r}\right)\right]^2}} \label{eq:A2}
\end{equation}

Here, \(w_i\) is the weight assigned to the i-th data point. The weight is computed as follows: For each data point, calculate the wavelength differences with its immediately preceding and succeeding neighbors, then average these differences (for the first and last points, use the differences with the single adjacent point). Normalize these average differences by dividing each by the sum of all such averages across the spectrum. Finally, divide the resulting normalized value by the \kamo’s reflectance measurement error of the corresponding (i-th) data point to obtain \(w_i\).

\subsection{\chadded{Additional Spectral Comparisons}} \label{ap:A2new}

\begin{figure*}[ht!]
    \plotone{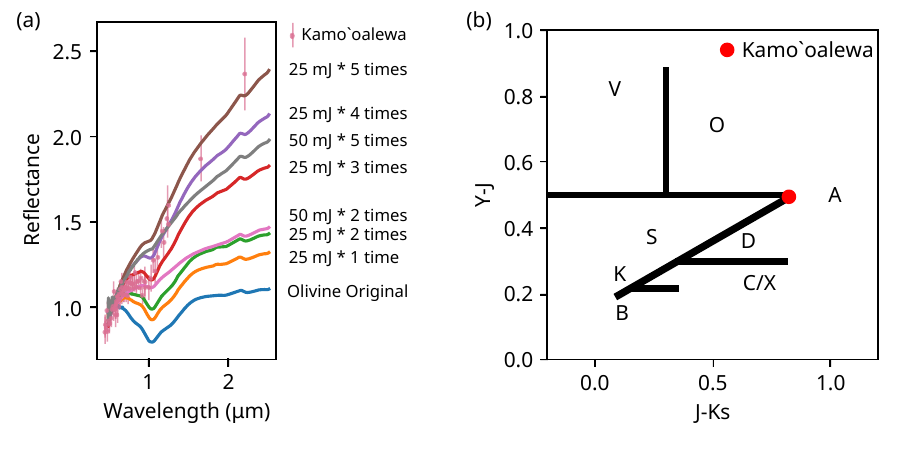}
    \caption{\chadded{Additional spectral comparisons of the \kamo\ spectrum. (a) Direct comparison with laser-irradiated olivine spectra \citep{2017A&A...597A..50Y}; (b) Comparison of NIR colors using the method of \citet{2018A&A...617A..12P}. In panel (a), the morphology of the 1-\um\ absorption band in the \kamo\ spectrum resembles that of the olivine spectrum after 3-4 irradiations with a 25 mJ laser. In panel (b), the data points for \kamo\ fall within the region associated with A-type asteroids (see Figure 1 in \citet{2018A&A...617A..12P} for details).}}
    \label{fig:newapa2}
\end{figure*}

\chadded{In addition to the quantitative CPRMS similarity analysis of \kamo\ with the mean spectra in the Bus-DeMeo taxonomy, we performed a simple direct comparison between the spectrum of \kamo\ and that of laser-irradiated olivine, as well as a comparative analysis of the NIR-band colors using only the J, H, and K data points based on the method proposed by \citet{2018A&A...617A..12P}. These comparisons are shown in Figure \ref{fig:newapa2}(a) and Figure \ref{fig:newapa2}(b), respectively. The morphology of the 1-\um\ absorption band in \kamo's spectrum resembles that observed in the olivine sample after 3-4 irradiations with a 25 mJ/pulse laser. In the NIR band, \kamo's data points in Figure \ref{fig:newapa2}(b) fall within the region associated with A-type asteroids (see Figure 1 in \citealp{2018A&A...617A..12P} for details).}

\chadded{These results support the high olivine content of \kamo. The former indicates that \kamo\ may exhibit a low degree of space weathering. The latter confirms that the H and K data points bias the spectral match toward the reddest template (A-type), which is not well supported by the spectrum at shorter wavelengths (S-type). This finding aligns with the results in Figure 1 and demonstrates that \kamo's spectrum exhibits characteristics of both S-type and A-type asteroids: S-type features in the range up to 1.25 \um\ and A-type features beyond 1.25 \um. The presence of both types is consistent with our hypothesis that \kamo's olivine content may exceed that of typical S-type asteroids.}

\subsection{Band Center Extraction} \label{ap:AA2}

The spectrum of \kamo\ exhibits a steep spectral slope and a prominent absorption feature at 1 \um, with limited data points available near the 2 \um\ absorption band \citep{2021ComEE...2..231S,sharkeydata2021}. In fact, this spectrum has only two data points beyond 1.25 \um\ (H and K bands).

A common method for analyzing the mineral composition of asteroids is to calculate the ratio of the 2 \um\ absorption area to the 1 \um\ absorption area (band area ratio, BAR). This ratio reflects the relative proportions of olivine and orthopyroxene (OPX), as described by \citet{1986JGR....9111641C} and \citet{2015aste.book...43R}. However, because \kamo's spectrum has particularly sparse data points near 2 \um, making it impossible to determine the 2 \um\ absorption area and thus calculate the BAR. An alternative approach is the method proposed by \citet{1993Icar..106..573G}, in which the x-axis represents the BAR of the target spectrum and the y-axis represents the band center (BC) of the 1 \um\ absorption feature (BC1). Although we cannot determine the x-axis value (BAR) as noted earlier, mixtures of olivine and orthopyroxene, along with observed spectra of S-type asteroids, have demonstrated a relatively monotonic correlation between BAR and BC1 \citep{1986JGR....9111641C,1993Icar..106..573G}. Therefore, we can infer Kamo'oalewa's potential olivine and OPX proportions solely from the calculated BC1.

To this end, we adopted the method of \citet{2014Icar..234..132H} to calculate the band center of \kamo's 1 \um\ absorption feature, utilizing only its VNIR spectrum up to 1.25 \um. Specifically, after removing the continuum from the spectrum \citep{2020Icar..35013901M}, we performed polynomial fitting on the central region within the 1 \um\ absorption feature to obtain the specific wavelength value of BC1. Due to the noticeable noise in \kamo's spectrum, we applied Savitzky-Golay filtering \citep{SGFM} with a window length of 11 and a polynomial order of 3.
As an integral part of our analysis, we calculated BC1 values for a series of spectra, including those of olivine and orthopyroxene mixtures prepared by McCoy (Tabel \ref{tab:mccoy}) from the RELAB database \citep{RELABdb}, as well as those of space-weathered olivine simulated via pulsed laser irradiation \citep{2017A&A...597A..50Y}. For each spectrum, we performed BC1 analysis to both the truncated portion up to 1.25 \um\ and the full wavelength range (without truncation),thereby validating our validate our approach for \kamo.

\begin{deluxetable*}{cc}
\tablewidth{0.8 \textwidth}
\tablecaption{McCoy Spectra IDs and Compositions} \label{tab:mccoy}
\tablehead{
\colhead{Spectrum ID} & \colhead{Composition (wt. \%)}
}
\startdata
c1ag08 & 100\% OLV \\
c1ag20 & 90\% OLV, 10\% OPX \\
c1ag14 & 70\% OLV, 30\% OPX \\
c1ag19 & 50\% OLV, 50\% OPX \\
c1ag18 & 30\% OLV, 70\% OPX \\
c1ag17 & 10\% OLV, 90\% OPX \\
c1ag09 & 100\% OPX \\
\enddata
\tablecomments{The spectral data presented in this table were obtained from the RELAB spectral database \citep{RELABdb}. The samples are owned by Tim McCoy. OLV: Olivine; OPX: Orthopyroxene.}
\end{deluxetable*}





\section{Scatter Plots Showing Space Weathering Characteristics} \label{ap:B}

The R950/R750 versus R750 scatter plot—where the ordinate represents the ratio of the reflectance at 950 nm to that at 750 nm, and the abscissa represents the reflectance at 750 nm—serves as a well-established method for depicting the degree and spectral trends of lunar-style space weathering \citep{1995Sci...268.1150L,2000JGR...10520377L}. In this scatter plot, the distance from a sample data point to the hypothetical point in the upper-left corner is used to quantify the degree of space weathering for the sample, while the angle between the abscissa and the line connecting the sample data point to this hypothetical point represents the space weathering spectral trend, which is related to the sample's FeO content. Studies have investigated the relationship between OMAT and Is/FeO values (e.g., \citealt{2024A&A...682A.112W}), offering additional insights into space weathering. We adapted this approach to accommodate the properties of normalized spectra by retaining the ordinate while modifying the abscissa to R750/R550. This adaptation enables seamless integration of absolute and normalized reflectance spectra. Spectral normalization within a specific band involves dividing the reflectance values across all bands by the reflectance in that band. For any given spectrum, the ratio of reflectances between two bands remains invariant, whether derived from absolute or normalized values, as the normalization factor cancels out in both numerator and denominator.

For brevity, we refer to the R950/R750 versus R750 plot as the traditional model and our adapted R950/R750 versus R750/R550 plot as the modified model. In the traditional model, the key wavelengths of 0.75 \um\ and 0.95 \um\ correspond approximately to the left shoulder and near the center of the 1 \um\ absorption feature in lunar surface spectra, respectively. Thus, R750 indicates the overall spectral brightness, while R950/R750 inversely correlates with the depth of the 1 \um\ absorption feature. As lunar-style space weathering intensifies, spectra darken (decreasing R750), and the 1 \um\ absorption feature shallows, leading to a relative increase in R950 and a corresponding rise in R950/R750. Consequently, data points in the traditional model shift from the lower right to the upper left with increasing weathering. Additionally, the inclination angle between the line connecting a data point to the theoretical space weathering saturation point and the horizontal line passing through the saturation point reflects the FeO content.

In the modified model, normalization largely eliminates brightness variations from the original absolute reflectance spectra. The ordinate retains its interpretation from the traditional model, whereas the abscissa now represents the spectral continuum slope. Accordingly, increasing space weathering steepens the spectral slope and reduces the 1 \um\ absorption depth, resulting in data points progressing from the lower left to the upper right—a clear and robust pattern in our model. The speculated progression from lower right to upper left, indicative of enhanced weathering efficiency, warrants further investigation. The traditional model exhibits two prominent diagonal patterns, while the modified model displays one dominant pattern and one latent pattern; the latter's attenuation likely stems from the loss of brightness information during normalization. However, this attenuation also offers an advantage by enabling the integration of a broader range of data in the analysis, as brightness variations among different asteroids, as well as between asteroids and the Moon, are primarily attributable to observational conditions rather than space weathering effects. Under these conditions, we eschew fitting lines of varying slopes to identify a weathering saturation point (in the traditional model) and instead employ the straight-line distance from each data point to the origin as a normalized optical maturity index, termed NOMAT. The calculation is given by Equation (\ref{eq:A3}):

\begin{equation}
NOMAT=\sqrt{\left(\frac{R750}{R550}\right)^2+\left(\frac{R950}{R750}\right)^2} \label{eq:A3}
\end{equation}

\noindent A larger NOMAT value thus indicates higher maturity.

\section{Validation Details of Is/FeO vs. NOMAT Model} \label{ap:C}

This appendix details the validation of the empirical relation between Is/FeO and NOMAT derived in this study. The validation incorporates typical Is/FeO values for CE-5 samples \citep{2024Icar..41015892Q}, the Hiroi space weathering model \citep{2006Natur.443...56H}, and the conversion relation between Is/FeO and \npfe\ proposed by \citep{1980LPSC...11.1697M}. These elements are discussed in the subsections below. First, we outline the general validation strategy. The validation comprises two independent components: (1) We utilize the spectra of CE-5 samples and their measured Is/FeO values to assess the model. (2) We estimate the Is/FeO value by fitting the \npfe\ vol.\% for \kamo\ and compare it with the Is/FeO value obtained by our empirical model relating Is/FeO to NOMAT. Notably, this second component also offers insights into the material composition of \kamo.

\subsection{Is/FeO Value for CE-5 Samples} \label{ap:c1}

The CE-5 samples were collected from the northern region of the Rümker Mountains on the Moon, exhibiting relatively uniform properties across specimens \citep{2022NSRev...9B.188L}. Based on the spectrum measured by \citet{2022NatCo..13.3119L}, we calculated the NOMAT value to be approximately 1.63. Substituting this into our empirical model yields an Is/FeO value range of 52.49±19.61 (1\sig\ prediction interval). The average Is/FeO value of the CE-5 samples, as measured by \citet{2024Icar..41015892Q}, using ferromagnetic resonance, is 14, which lies outside the 1\sig\ prediction interval of our model but within its 2\sig\ prediction interval, indicating its sufficient accuracy.

For lunar samples, Is/FeO values of $0-30$ indicate an immature state, values of $30-60$ indicate a submature state, and values greater than 60 indicate a mature state \citep{1976LPSC....7..315M,1978LPSC....9.2287M}.

\subsection{Hiroi Space Weathering Model} \label{ap:c2}

The Hiroi space weathering model, proposed by \citet{2006Natur.443...56H} based on spectral data from the Hayabusa spacecraft's in situ observations of the S-type near-Earth asteroid (NEA) Itokawa, estimates the volume fraction of \npfe\ on planetary surfaces. The core equation of this model is:

\begin{equation}
    ln{R}\left(\lambda\right)\cong-\left\{\alpha_\mathrm{h}\left(\lambda\right)+\frac{36\pi}{\lambda}\varphi z\left(\lambda\right)\right\}d_\mathrm{e} \label{eq:a4}
\end{equation}

According to \citet{2006Natur.443...56H}, \(R\left(\lambda\right)\) represents the absolute reflectance, \(\alpha_\mathrm{h}\) is the absorption coefficient of the asteroid host material, \vphi\ is the volume fraction of \npfe\ within the host material, \(d_\mathrm{e}\) is the mean optical path length (MOPL), and \(z\left(\lambda\right)\) is calculated using the following formula:

\begin{equation}
    z\left(\lambda\right)=\frac{n_\mathrm{h}^3n_{\mathrm{Fe}}k_{\mathrm{Fe}}}{\left(n_{\mathrm{Fe}}^2-k_{\mathrm{Fe}}^2+2n_\mathrm{h}^2\right)^2+\left(2n_{\mathrm{Fe}}k_{\mathrm{Fe}}\right)^2} \label{eq:a5}
\end{equation}

\noindent In this equation, \(n_\mathrm{h}\)
is the real part of the refractive index of the host material, while \(n_\mathrm{Fe}\) and \(k_\mathrm{Fe}\) are the real and imaginary parts of the refractive index of elemental iron, respectively.

The host material refers to the asteroid's composition unaffected by space weathering, which is reduced to \npFe\ upon exposure. For granular samples, MOPL corresponds to the effective particle diameter, typically the average of the maximum and minimum particle sizes. For rocks or surfaces with insufficient particles, MOPL estimation becomes more complex and may depend on their physical properties such as rock porosity.

To apply this model for estimating the \npFe\ volume fraction on an asteroid surface, the measured reflectance spectrum of the asteroid is required. A series of \vphi\ and {\(d_\mathrm{e}\)} values are substituted into Equation (\ref{eq:a4}) to generate fitted R values. The \vphi\ and  \(d_\mathrm{e}\) that minimize the difference between fitted and measured reflectance are considered the surface values. Additionally, the model requires offsetting \(\ln(R)\) to zero at 1.25 \um\ at any time (note the distinction from normalization: normalization scales all wavelengths by a constant multiplier or divisor, whereas offsetting adds or subtracts a constant from all reflectance values).

Evidently, the fitting process requires two optical parameters of the asteroid: \(\alpha_\mathrm{h}\) and \(n_\mathrm{h}\). The Hiroi model addresses this by selecting an analog material with a composition similar to the asteroid but exhibiting minimal space weathering. For Itokawa, \citet{2006Natur.443...56H} used LL5 ordinary chondrite (OC) Alta'ameem as the analog. By measuring the spectrum of the analog's powdered sample in the laboratory, using the effective particle diameter as \(d_\mathrm{e}\), and approximating \vphi\ to zero (due to low space weathering in the selected analog), \(\alpha_\mathrm{h}\) is calculated from Equation (\ref{eq:a4}) and approximated for the asteroid. The value of \(n_\mathrm{h}\) is estimated from the analog's mineral proportions.

A key advantage of the Hiroi model is its applicability to asteroids like \kamo, which have only normalized reflectance spectra. This stems from the requirement to offset \(\ln(R)\) to zero at a specific wavelength. For a spectrum normalized at any band, the normalizing operation equates to subtracting a constant from the logarithm of the original spectrum, which is nullified during offsetting due to logarithmic properties.

\begin{figure*}[ht!]
    \plotone{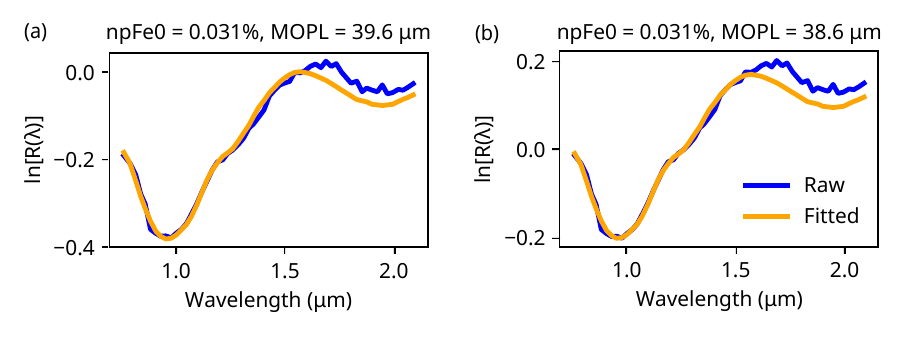}
    \caption{Schematic diagram comparing results before and after modifications to the Hiroi model, using spectra from the bright and blue areas of Itokawa as presented in \citet{2006Natur.443...56H}. (a) Fitting results from Hiroi's original model (absolute reflectance spectrum, offset to zero ordinate at 1.54 \um; spectral similarity calculation range: 0.76–1 \um). (b) Fitting results from the fine-tuned model (spectrum normalized at 1.2 \um, offset to zero ordinate at 1.25 \um; spectral similarity calculation range: 0.76–1.25 \um). The unmodified Hiroi model reproduces the results from \citet{2006Natur.443...56H}. The \npFe\ volume fraction from the modified model is consistent with that from the unmodified model, while the MOPL length differs by only 1 \um. These adjustments (using normalized spectra, altering the offset band, and expanding the spectral similarity calculation range) do not impact model performance significantly.}
    \label{fig:A1}
\end{figure*}

We implemented minor modifications to the Hiroi model to better suit the spectrum of \kamo: (1) absolute reflectance was replaced with relative (normalized) reflectance during fitting; (2) the offset wavelength was changed from 1.54 \um\ to 1.25 \um\ (as \kamo's spectrum lacks data at 1.54 \um, where interpolation would introduce significant errors); and (3) the primary spectral similarity assessment interval was expanded from 0.76–1 \um\ to 0.6–1.25 \um\ (to accommodate the broader wavelength range of \kamo's spectrum compared to Itokawa's). Figure \ref{fig:A1} shows the fitted spectrum of Itokawa's bright area \citep{2006Natur.443...56H} using both the original and modified models. The fitted \vphi\ remains unchanged, while MOPL varies by only 1 \um, confirming that these modifications have negligible impact on model performance.

Note that the modified Hiroi model demonstrated in Figure A1 still differs slightly from the version applied to \kamo. This discrepancy arises from the differing wavelength ranges of the spectra for \kamo\ and Itokawa. In the model applied to \kamo, the spectrum is normalized at 0.55 \um, with the spectral similarity calculation range set to 0.6–1.25 \um. However, the shortest wavelength in Itokawa's spectrum is 0.76 \um, rendering outward interpolation inadvisable. Nonetheless, the verification presented above confirms that the three key modifications to the model—using normalized spectra, altering the offset band, and expanding the spectral similarity calculation range—do not significantly affect its performance.

When applying the modified Hiroi model to \kamo, selecting an appropriate host material analog is crucial. Given the two hypotheses for \kamo's origin—main belt or the Moon \citep{2021ComEE...2..231S,2024LPICo3040.1845Z}—we considered both, subdividing the lunar hypothesis into mare and highland origins. Accordingly, we selected three analogs with low space weathering: LL5 OC Alta'ameem, lunar mare sample LSCC 12030 (particle size 20–45 \um) \citep{2001JGR...10627985T}, and lunar highland sample 67711 (particle size 0–250 \um) \citep{2012Icar..220...51B}. The \(n_\mathrm{h}\) values for the lunar mare and highland analogs were estimated based on mineral ratios from \citep{2001JGR...10627985T} and \citep{2019GeCoA.266...17T}, combined with mineral refractive indices from \citet{2011JGRE..116.9001L}, yielding 1.62 and 1.59, respectively; for the OC analog, we adopted the value from \citet{2006Natur.443...56H}. The \(n_\mathrm{Fe}\) and \(k_\mathrm{Fe}\) we use are from \citet{1974PhRvB...9.5056J}.

\begin{figure*}
    \centering
    \plotone{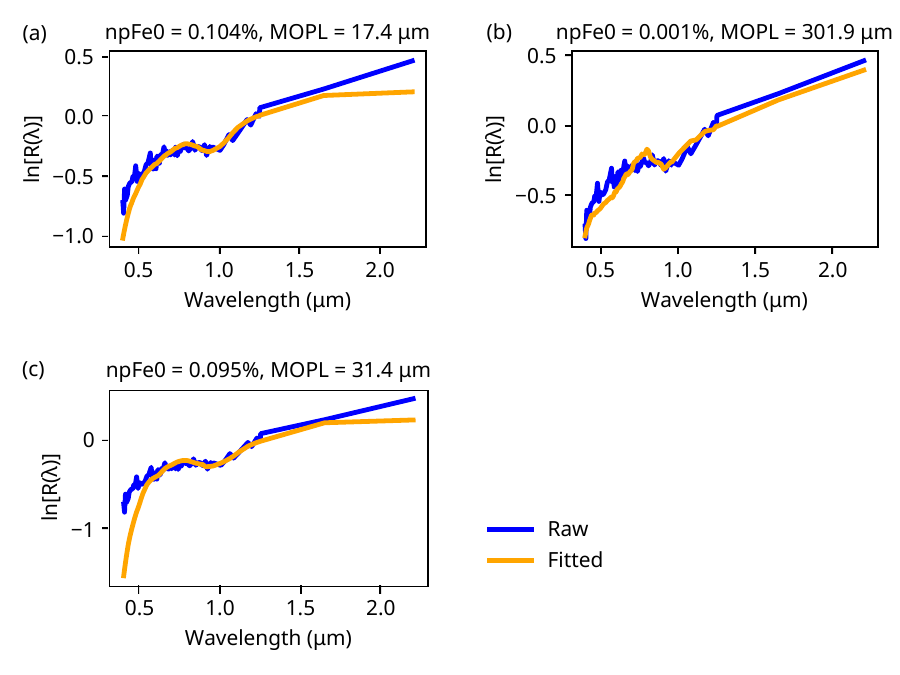}
    \caption{Fitting results for the volume fraction of \npFe\ and MOPL in the spectrum of \kamo\ using the modified Hiroi model. (a) Results using lunar mare sample 12030 (20-45 \um) as the analog for \kamo. (b) Results using lunar highland sample 67711 as the analog. (c) Results using LL5 ordinary chondrite Alta'ameem as the analog.}
    \label{fig:A2}
\end{figure*}

For the OC, lunar mare, and highland analogs, the fitted \vphi\ values are 0.095, 0.104, and 0.001 vol.\%, with corresponding MOPL values of 31.4, 17.4, and 301.9 \um. The fitting results are illustrated in Figure \ref{fig:A2}. Notably, under the OC and lunar mare analogs (corresponding to main-belt and lunar mare origins), the MOPL values are comparable, and \vphi\ values are both relatively high. In contrast, the highland analog (lunar highland origin) yields an excessively large MOPL and low \vphi, with Figure \ref{fig:A2}(b) showing poor spectral shape agreement between fitted and measured data in the interested 0.6-1.25 \um\ region. This suggests that \kamo\ is unlikely to originate from lunar highlands, but rather from the main belt or lunar mare regions. However, as determining \kamo's origin is not the objective of this study, we discuss it only briefly here; in the main text and subsequent sections, we focus on the main-belt and lunar mare origin scenarios.

\subsection{Morris Is/FeO and \npfe\ Empirical Model} \label{ap:c3}

\citet{1980LPSC...11.1697M} proposed an empirical relation between the weight percent (wt.\%) of \npFe\ and the Is/FeO value for lunar samples:

\begin{equation}
\psi_{npFe0}=\left(3.2\times{10}^{-4}\right)\left(\psi_{FeO}\right)\left(Is/FeO\right) \label{eq:a6}
\end{equation}

In this equation, \(\psi_{FeO}\) is the wt.\% of FeO, and \(\psi_{npFe0}\) is the wt.\% of \npfe. We employ this relation to estimate the Is/FeO value of \kamo\ based on the \npFe\ volume fraction fitted using the Hiroi model. This method provides an independent estimate of Is/FeO for \kamo, thereby serving to validate the accuracy of the Is/FeO value derived from our empirical Is/FeO versus NOMAT relation.

Following the discussion in the previous subsection, we applied the Morris relation under two scenarios: \kamo\ originating from the main belt or from lunar mare regions, as these yield different \npFe\ contents. To utilize Equation (\ref{eq:a6}), the \npFe\ volume fraction must first be converted to a mass fraction using the formula:

\begin{equation}
\psi_{npFe0}=\varphi\times\frac{\rho_{Fe}}{\rho_m} \label{eq:A7}
\end{equation}



\noindent Here, \vphi\ is the vol.\% of \npfe\ (the same as in Equation \ref{eq:a4}). \(\rho_{Fe}\) is the density of \npFe, and \(\rho_m\) is the bulk density of the host material containing \npFe.

Substituting Equation (\ref{eq:A7}) into Equation (\ref{eq:a6}) yields:

\begin{equation}
Is/FeO=\frac{\varphi\times\rho_{Fe}}{\left(3.2\times{10}^{-4}\right) \times \rho_m \times \psi_{FeO}}
\label{eq:a8}
\end{equation}



For this calculation, we adopted \(\rho_{Fe}\) as the standard density of iron (7874 \(\mathrm{kg/m^{3}}\)), with \(\rho_m\) values of 3327 and 3270 \(\mathrm{kg/m^{3}}\)) for typical ordinary chondrites (OC) \citep{2003M&PS...38.1533W} and lunar mare samples \citep{2012GeoRL..39.7201K}, respectively. Similarly, the FeO mass fractions were taken as 16.0 and 17.19 wt.\%\ for High-FeO OC \citep{1995GeCoA..59.4951K} and lunar mare materials \citep{2000JGR...10520333E}. The resulting Is/FeO values are 43.92 and 45.53, respectively, both of which fall within the 1\sig\ prediction interval of our model predictions.



\bibliography{Liu_refs}{}
\bibliographystyle{aasjournalv7}



\end{document}